\documentclass[prb,aps,twocolumn,floats,superscriptaddress]{revtex4} 

\usepackage{ulem}
\usepackage{amsmath}
\usepackage{graphicx}
\usepackage{epsfig}
\usepackage{psfrag}
\usepackage{mathdots}
\usepackage{color}
\newlength{\upit}\upit=0.1truein
 
\newcommand{\ltappr}{{{\lower4pthbox{$<$} } \atop \widetilde{ \ \ \ }}} 
\newlength{\bxwidth}\bxwidth=1.5 truein

\begin{document} 
\newcommand{\dg}{^{\dagger }} 
\newcommand{\vk}{\vec k} 
\newcommand{\vq}{{\vec{q}}} 
\newcommand{\vp}{\bf{p}} 
\newcommand{\al}{\alpha} 
\newcommand{\be}{\beta} 
\newcommand{\si}{\sigma} 
\newcommand{\rarrow}{\rightarrow}
\newcommand{\zmatrix}[4]{\left[\!\begin{matrix}#1 & #2\cr #3&#4\end{matrix}\!\right]}
\newcommand{\vertvec}[2]{\left(\!\!\begin{array}{c}#1\cr#2\end{array}\!\!\right)} 
\newcommand{\vertvecT}[3]{\left(\!\begin{array}{c}#1\cr#2\cr#3\end{array}\!\right)} 
\def\fig#1#2{\includegraphics[height=#1]{#2}} 
\def\figx#1#2{\includegraphics[width=#1]{#2}} 
\newlength{\figwidth} 
\figwidth=8.6cm 
\newlength{\shift} 
\shift=-0.2cm 
\newcommand{\fg}[3] 
{ 
\begin{figure}[ht] 
 
\vspace*{-0cm} 
\[ 
\includegraphics[width=\figwidth]{#1} 
\] 
\vskip -0.2cm 
\caption{\label{#2} 
\small#3 
} 
\end{figure}} 
\newcommand{\fgb}[3] 
{ 
\begin{figure}[b] 
\vskip 0.0cm 
\begin{equation}\label{} 
\includegraphics[width=\figwidth]{#1} 
\end{equation} 
\vskip -0.2cm 
\caption{\label{#2} 
\small#3 
} 
\end{figure}} 
%\psdraft 
 
\newcommand \bea {\begin{eqnarray} } 
\newcommand \eea {\end{eqnarray}} 
\newcommand{\bk}{{\bf{k}}} 
\newcommand{\bx}{{\bf{x}}}
\newcommand{\bR}{{\bf{R}}}
\newcommand{\br}{{\bf{r}}}
\newcommand{\bq}{{\bf{q}}}
\newcommand{\e}{{\mathrm{e}}}
\newcommand{\new}[1]{\color{blue}#1 \color{black}}
\newcommand{\cut}[1]{\color{red}\sout{#1} \color{black}}
\newcommand{\sgn}{{\rm sgn}}

\title{The symplectic-$N$ t-J model and s$_\pm$ superconductors}

\author{Rebecca Flint}
\affiliation{
Center for Materials Theory, Department of Physics and Astronomy,
Rutgers University,\\
136 Frelinghuysen Rd., Piscataway, NJ 08854-8019, USA}
\affiliation{
Department of Physics, Massachusetts Institute for Technology,\\
77 Massachusetts Avenue, Cambridge, MA 02139-4307,USA}
\author{Piers Coleman}
\affiliation{
Center for Materials Theory, Department of Physics and Astronomy,
Rutgers University,\\
136 Frelinghuysen Rd., Piscataway, NJ 08854-8019, USA}
\affiliation{
Department of Physics, Royal Holloway, University
of London,\\
Egham, Surrey TW20 0EX, UK.
}

\begin{abstract}
% should soften - not 100% clear that it's $s_\pm$
The possible discovery of $s_\pm$ superconducting gaps in the moderately correlated iron-based superconductors has raised the question of how to properly treat $s_\pm$ gaps in strongly correlated superconductors.  Unlike the d-wave cuprates, the Coulomb repulsion does not vanish by symmetry, and a careful treatment is essential.  Thus far, only the weak correlation approaches have included this Coulomb pseudopotential, so here we introduce a symplectic $N$ treatment of the $t-J$ model that incorporates the strong Coulomb repulsion through the complete elimination of on-site pairing. 
Through a proper extension of time-reversal symmetry to the large $N$ limit, symplectic-$N$ is the first superconducting large $N$ solution of the $t-J$ model.  For d-wave superconductors, the previous uncontrolled mean field solutions are reproduced, while for $s_\pm$ superconductors, the $SU(2)$ constraint enforcing single occupancy acts as a \emph{pair chemical potential} adjusting the location of the gap nodes.  This adjustment can capture the wide variety of gaps proposed for the iron based superconductors: line and point nodes, as well as two different, but related full gaps on different Fermi surfaces.
\end{abstract}

\maketitle

\section{Introduction}

The new family of iron-based superconductors\cite{hosano08} has
expanded the study of high temperature superconductors from the single
band, d-wave cuprate superconductors to include multi-band
superconductors with full gaps.  Experimental\cite{lynn09} and
numerical\cite{si08,haule08,kutepov10} work suggest a range of
correlation strengths between different materials. From the
theoretical point of view, weak and strong correlation approaches
converge on many of the major features: most importantly, the predominantly $s_\pm$
structure of the superconducting
gap\cite{mazin08,chubukov08,si08,seo08,parish08,maier09,wang09,sknepnek09,yao09,hu11,hirschfeld11}.
The real materials are likely in the regime of moderate correlations
where both approaches are useful.  Unlike the d-wave cuprates, where
the strong Coulomb repulsion is eliminated by symmetry, these
multi-band $s_\pm$ superconductors require a careful treatment of the
Coulomb pseudopotential\cite{morel62}.  While this has been
incorporated into weak coupling
approaches\cite{chubukov08,maier09,sknepnek09,mazin09}, it has yet to
be included in strong correlation treatments based on the $t-J$ model.
With this in mind, we introduce the symplectic-$N$ $t-J$ model.  The
use of a large $N$ limit based on the symplectic group, $SP(N)$ allows
a proper treatment of time-reversal in the large-$N$
limit\cite{flint08,flint09,flint11}, making this the first
superconducting large $N$ treatment of the $t-J$ model.
Symplectic-$N$ also replaces the usual $U(1)$ constraint of
single-occupancy with a $SU(2)$ constraint that strictly eliminates
on-site pairing\cite{flint11}.  This $SU(2)$ constraint is essential
to the treatment of $s_\pm$ superconductors, where it acts as a
\emph{pair chemical potential}, adjusting the gap nodes to eliminate
the Coulomb repulsion.

This paper is intended as an introduction to the symplectic-$N$ $t-J$
model, illustrating the importance of the additional constraint with
the example of $s_\pm$ superconductors, and showing how this model
contains a range of gap behaviors reproducing different iron-based
materials.  We begin by reviewing the Coulomb pseudopotential in
section \ref{coulomb} and demonstrate the lack of superconductivity in
the $SU(N)$ $t-J$ model corresponding to the usual slave boson mean
field theory.  In section \ref{symplectic_tJ}, we introduce symplectic
Hubbard operators, which allow us to develop a superconducting
large-$N$ treatment of the $t-J$ model.  We demonstrate this mean
field theory on several examples in section \ref{examples}, before
discussing the range of possible future directions in
\ref{discussion}. 

\section{The Coulomb pseudopotential and the $t-J$ model}
\label{coulomb}

On-site pairing is disfavored by the Coulomb pseudopotential, which will cost a bare energy, $U N(0)$, the average of the Coulomb repulsion, $V(\br_i - \br_j) = e^2/|\br_i - \br_j|$ over the Fermi sea.  However, in the weak coupling limit, where we assume the pairing is mediated by the exchange of a boson with characteristic frequency $\omega_B$, the time scale of the pairing, $\omega_B$ is much longer than that of the Coulomb repulsion.  In other words, while the effective electron-electron interaction is attractive, it is also retarded, meaning the electrons like to be in the same place, but at different times, while the Coulomb repulsion is a nearly instantaneous repulsion of two electrons at the same place \emph{and} time.  The Coulomb pseudopotential is therefore renormalized\cite{morel62},
\begin{equation}
\mu^* = \frac{N(0)U}{1+N(0) U \log \frac{E_F}{\omega_B}}
\end{equation}
to weak coupling.  If $T_c \propto \omega_B \exp(-1/\lambda)$, the attractive interaction is replaced by $\lambda \rarrow \lambda - \mu^*$, which reduces $T_c$ slightly at weak coupling, but does not destroy superconductivity.  In BCS superconductivity, the bosons exchanged are phonons, and the Debye frequency, $\omega_D \ll E_F$.  However, in more strongly correlated superconductors, the two time scales are of the same order, and the Coulomb pseudopotential can drastically affect the superconductivity.  Strongly correlated examples, like the cuprate and heavy fermion superconductors, avoid this problem by developing a d-wave gap, where the pairing with a positive gap is exactly cancelled out by that with a negative gap, as guaranteed by the d-wave symmetry.  This choice of gap neutralizes the Coulomb pseudopotential.  However, the iron-based superconductors are widely believed to have an $s_\pm$ gap, where the amount of cancellation between positive and negative gap regions is not protected by symmetry, and depends strongly on the Fermi surfaces.  When this cancellation is incomplete, $\mu^*$ reduces $T_c$ and it is extremely important to consider this effect when mapping out the phase diagram, as it affects the relative stability of s- and d-wave gaps.  These effects have been incorporated in the weakly correlated solutions\cite{chubukov08,maier09,sknepnek09,mazin09}, but not yet in the strongly correlated approaches.

\fg{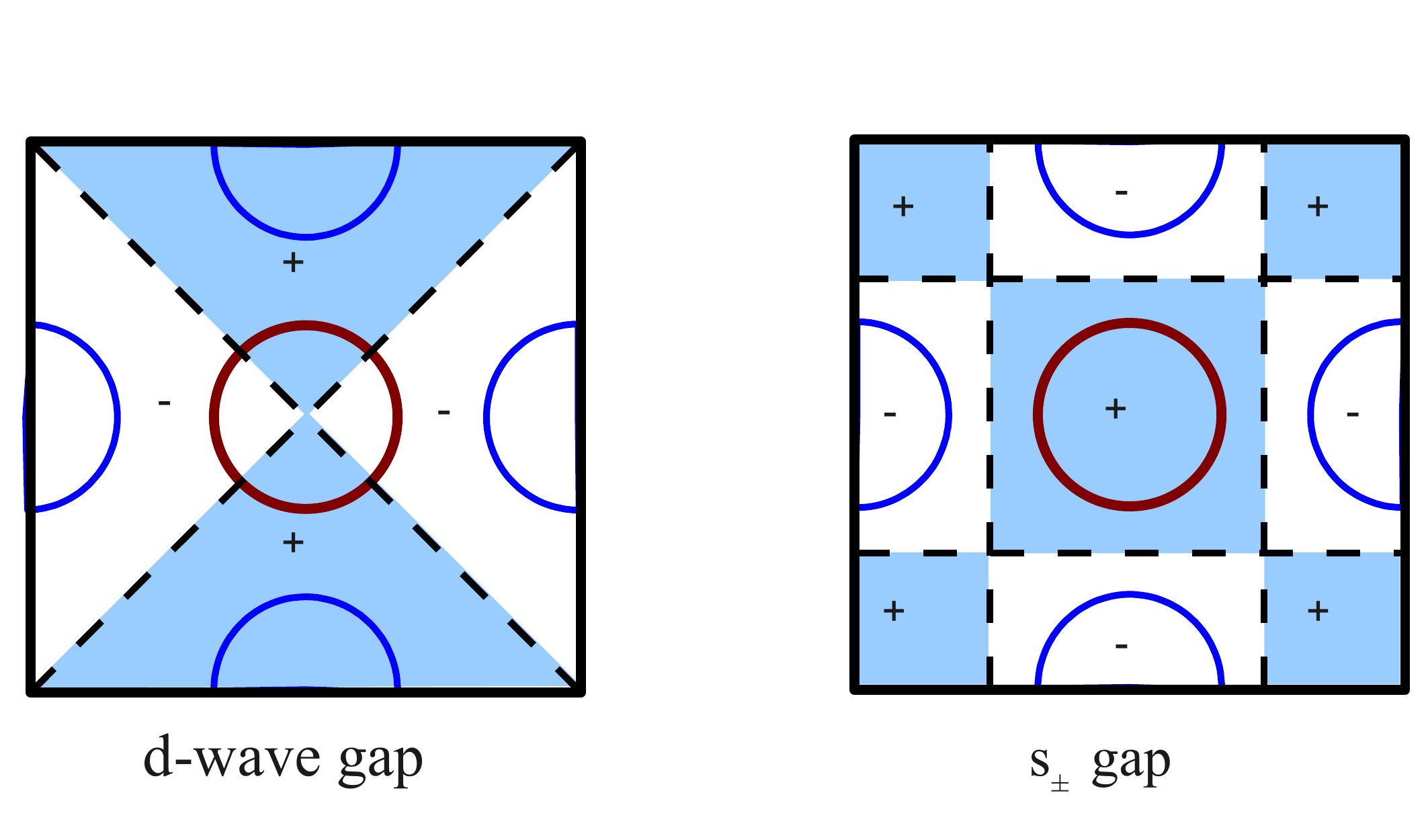}{pseudo}{(Left) In the d-wave gap, the cancellation of the superconducting order parameter over the Fermi surface is guaranteed by symmetry, as the positive (blue) regions will exactly cancel the negative (white) regions.  (Right) However, in the $s_\pm$ superconducting gap, the amount of cancellation is extremely sensitive to the Fermi surface.}

Here, we take the strongly correlated limit, $U \rarrow \infty$ to eliminate double occupancy, which corresponds to taking $\mu^* \rarrow \infty$.
The Heisenberg model describes the insulating half-filled limit of the $t-J$ model, but generally holes ($n <1$) or electrons ($n >1$) will hop around in an antiferromagnetic background.  Doubly occupied states must be avoided, and the hopping is \emph{not} that of free electrons.  Rather, it is \emph{projected hopping}, described by the $t-J$ model\cite{spalek77, anderson87},
\begin{equation}
H = - \sum_{ij} t_{ij} \left[ X_{\sigma 0}(i)X_{0\sigma}(j) +\mathrm{H.c.}\right] + \sum_{ij} J_{ij} \vec{S}_i \cdot \vec{S}_j.
\end{equation}
The Hubbard operators, $X_{ab} = |a\rangle \langle b|$, where
$|a\rangle = |0\rangle, |\sigma\rangle$ ensure that only empty sites,
or holes can hop (or for $n >1$ that electrons can only hop from
doubly occupied sites to singly occupied sites).  Here, $X_{\sigma0}$
are projected hopping operators
%\cut{, and they satisfy the anti-commutation relations,}
%\begin{equation}
%\cut{\{X_{\sigma 0},X_{0\sigma'} \} = X_{\sigma \sigma'} + X_{00} \delta_{\sigma \sigma'},}
%\end{equation}
%\cut{where the spin, $S_{\sigma \sigma'}$ is the traceless form of
%the Hubbard operator, $S_{\sigma \sigma'} = X_{\sigma \sigma'}-\frac{\delta_{\sigma \sigma'}}{2} X_{\tau \tau}$}, while
%$X_{00}$ is a projection operator into the empty state.

Exact solutions of the $t-J$ model are unavailable, and the typical approach is to write down a mean field solution using the slave boson approach\cite{kotliarliu88,wenlee96,lee06}, which divides the electron into charged, but spinless holons and neutral spinons.  The most common choice is the $U(1)$ slave boson representation\cite{coleman83},
\begin{equation}
X_{\sigma 0} = f\dg_\sigma b,
\end{equation}
so-called because it is invariant under $U(1)$ gauge transformations.
However, mean field solutions do not necessarily satisfy all the
conditions on the full model, and may not maintain the $\mu^* \rarrow
\infty$ limit.  Large $N$ approaches generate mean-field solutions by
extending the $SU(2)$ $t-J$ model to some larger group.  When
developing a large $N$ treatment of the hopping term, one must take
care that the two terms are consistent, or in other words, that the
charge fluctuations described by the $t$ term generate the spin
fluctuations in the Heisenberg term. 
The algebra of these operators, given by
\begin{equation}
\{X_{\alpha 0},X_{0\beta} \} = f\dg_\alpha f_\beta + b\dg b \delta_{\alpha \beta}.
\end{equation}
extends the algebra of Hubbard operators from $SU (2)$ to $SU (N)$,
and we see that two charge fluctuations in sequence give rise to 
a spin fluctuation described by the $SU (N)$ spin operator
$X_{\alpha \beta }= f\dg_{\alpha}f_{\beta}$.

The large $N$ limit of the full $SU(N)$ $t-J$ model is:
\begin{equation}
\label{largeNtJ}
H = -\!\sum_{\langle ij\rangle}\! \frac{t_{ij}}{N}\! f\dg_{i\alpha} b_i b_j\dg f_{j\alpha} + \!\sum_{ij}\!\frac{J_{ij}}{N}\! \left(f\dg_{i\alpha} f_{j\alpha}\!\right)\! \left(f\dg_{j\beta} f_{i\beta}\!\right)\!.
\end{equation}
Decoupling the $J$ term yields a dispersion for the spinons, but not pairing\cite{affleckmarston88}.  There is no superconductivity in this large $N$ limit.  

\section{The symplectic-$N$ $t-J$ model}
\label{symplectic_tJ}

A superconducting large $N$ limit requires a proper definition of time-reversal, as Cooper pairs can only form between time-reversed pairs of electrons.  The inversion of spins under time-reversal is equivalent to symplectic symmetry, and the only way to preserve time-reversal in the large $N$ limit is to use symplectic spins\cite{flint08,flint09},
\begin{equation}
S_{\alpha \beta} = f\dg_\alpha f_\beta - \tilde{\alpha}\tilde{\beta} f\dg_{-\beta} f_{-\alpha},
\end{equation}
where $\alpha$ ranges from $-N/2$ to $N/2$ and $\tilde{\alpha} =
\mathrm{sgn}(\alpha)$.
Here we use the fermionic representation because we are interested in
the doped spin liquid states that become superconductors.  Introducing
doping means introducing a small number of mobile empty states.  When
an electron hops on and off a site, it can flip the spin of the site.
Mathematically, this implies that the anticommutator of two
Hubbard operators generates a spin operator. 
In a symplectic-$N$ generalization
of the t-J model, anticommuting two such Hubbard operators must
generate a symplectic spin, satisfying the relations:
\begin{eqnarray}
\label{algebra}
\left\{X_{\alpha 0},X_{0\beta }\right\} &=& X_{\alpha\beta} + X_{00} \delta_{\alpha\beta} \\
&=& S_{\alpha \beta} + \left(X_{00}+\frac{X_{\gamma
\gamma}}{N}\right)\delta_{\alpha,\beta},\nonumber
\end{eqnarray}
where the last equality follows from the traceless definition of the
symplectic spin operator, $S_{\alpha \beta} = X_{\alpha \beta} -
\frac{X_{\gamma \gamma}}{N}\delta_{\alpha\beta}$. 
When we represent
the Hubbard operators with slave bosons, the symplectic projected
creation operators take the following form\cite{flint11},
\begin{equation}
X_{\alpha 0} = f\dg_\sigma b + \tilde{\sigma} f_{-\sigma} a,
\end{equation}
so that the other two Hubbard operators take the form
\begin{eqnarray}\label{l}
X_{\alpha \beta }&=& S_{\alpha \beta }+\delta_{\alpha \beta }\cr
X_{00}&=& b\dg b+a\dg a.
\end{eqnarray}
 This double slave boson
form for
Hubbard operators was derived by Wen and Lee\cite{wenlee96}
as a way of extending the local  $SU(2)$ symmetry of spin to
include charge fluctuations. 
In our approach the $SU(2)$ symmetry
appears as a consequence of the time-inversion properties of
symplectic spins for all even $N$, which permits us to carry out a
large $N$ expansion.  The Nambu notation, $B\dg = (b\dg,
a\dg)$ and $\tilde{f}\dg = (f\dg_\alpha, \tilde{\alpha} f_{-\alpha})$
simplifies the expressions, as $X_{\alpha 0} = \tilde{f}\dg_\alpha B$
and the hopping term of symplectic-$N$ $t-J$ model can be written,
\bea
H\!\! & = &\!\!\!-\!\sum_{ij}\!\frac{t_{ij}}{N}\!\!\left[\!\left(f\dg_{i \alpha}b_i + \tilde{\alpha}f_{i -\alpha} a_i\right)\!\!\left(f_{j \alpha}b_j\dg + \tilde{\alpha}f\dg_{j -\alpha} a_j\dg\right)\! +\! \mathrm{H.c.}\!\right]\cr
& = & \!\!\!-\!\sum_{ij}\!\frac{t_{ij}}{N}\left( \tilde{f}\dg_{i\alpha} B_i B_j\dg \tilde{f}_{j\alpha} + \mathrm{H.c.}\right)
\eea

{Restricting the spin and charge fluctuations to the physical subspace
requires that we fix the Casimir of the Hubbard operators\cite{hopkinson},
\begin{equation}\label{}
{\cal C}= \vec{S}_j^2 +\left[X_{0\alpha }, X_{\alpha
0}\right]- (X_{00}+1)^{2},
\end{equation}
where $\vec{S}_{j}^{2}= \frac{1}{2}\sum_{\alpha \beta }S_{\alpha \beta } (j) S_{\beta
\alpha } (j)$.
A detailed calculation (see Appendix) shows that 
\begin{equation}\label{}
{\cal C} = (N/2)^{2}-1 - \vec{\Psi}_j^2 ,
\end{equation}
where $\vec \Psi \equiv (\Psi\dg,\Psi,\Psi_{3})$ is given by
\begin{eqnarray}\label{l}
\Psi_{3}&=& n_{f}+ n_{b}-n_{a}- \frac{N}{2}\cr
\Psi\dg & = &
\sum_{\alpha >0}f\dg_{\alpha }f\dg_{-\alpha }
+ b\dg a\cr
\Psi & = & \sum_{\alpha >0}f_{-\alpha }f_{\alpha }+a\dg  b.
\end{eqnarray}
} In the infinite-$U$ limit, 
the Casimir, $\mathcal{C}$ is set to its maximal value, and we obtain the constraint $\vec{\Psi_j}=0$.
Writing out the condition that $\vec{\Psi_{j}}$
vanishes, we obtain
\bea
b_j\dg b_j - a_j\dg a_j + f\dg_{j\alpha} f_{j\alpha} = N/2 \cr
b_j\dg a_j +\sum_{\alpha >0}
\tilde{\alpha}f\dg_{j\alpha} f\dg_{j-\alpha}= 0 \cr 
a_j\dg b_j + \tilde{\alpha}f_{j-\alpha} f_{j\alpha}= 0.
\eea 
The first equation imposes the constraint on no double occupancy.
The second terms play the role of a Coulomb pair
pseudo-potential, forcing the net s-wave wave pair amplitude to 
be zero when superconductivity develops.
Under the occupancy constraint, there is only a single physical empty state,
which is
\begin{equation}
|0\rangle = \left(b\dg + a\dg \tilde{\alpha}f\dg_{-\alpha}f\dg_{\alpha}\right)|\Omega\rangle,
\end{equation}
for $N=2$.   
The physical interpretation of these terms becomes clearer
if we pick a particular gauge. Since we only have two flavors of
bosons and N flavors of fermions, the only way the bosons
contribute in the large N limit is by condensing.  As the bosons are
condensed at all temperatures, Fermi liquids and superconductors are
the only possible states; while this situation is clearly unphysical,
and will be resolved with 1/N corrections, it allows us to fix the
gauge in a particularly simple way by setting $a=0$ and condensing
only the $b$ bosons, $b^2 = Nx/2$ because the bosons
carry all the charge in the system.  The factor of $N/2$
makes the doping extensive in $N$.  The constraint simplifies to, 
\bea
f\dg_{j\alpha} f_{j\alpha} & = & \frac{N(1-x)}{2} \cr
\tilde{\alpha}f\dg_{j\alpha} f\dg_{j-\alpha} & = & 0 \cr
\tilde{\alpha}f_{j-\alpha} f_{j\alpha} & = & 0, \eea
In a mean field theory, these three constraints are
enforced
by a trio of Lagrange multipliers $\vec{\lambda} = (\lambda_+,
\lambda_-,\lambda_3)$ in a  constraint  term that takes the form
\begin{equation}\label{}
H_{C} = \sum_{j} \lambda_{3} \left[f\dg_{j\alpha}
f_{j\alpha}-\frac{N(1-x)}{2}\right]+
\lambda_{+} (\tilde{\alpha}f\dg_{j\alpha} f\dg_{j-\alpha})+ {\rm H.c}
\end{equation}
The first constraint is  clearly recognizable as
imposing Luttinger's theorem.  This term is present in the
conventional $U (1)$ slave boson approach\cite{kotliarliu88}.  The second terms impose severe constraints on the pair wavefunction when superconductivity develops, implementing the infinite Coulomb pseudopotential.
For d-wave superconductors
like the cuprates, which have been the main focus of previous $t-J$
model studies, these constraints are satisfied automatically, and at
the mean-field level, there is no difference between the
symplectic-$N$ limit and many of the previously considered
uncontrolled mean field theories\cite{kotliarliu88, wenlee96,vojta99}.  
However, for $s_\pm$ pairing,
these
additional constraints enforce the Coulomb pseudopotential, $\mu^*$
and have a large effect on the stability of $s_\pm$
superconductivity.

Once the bosons are condensed, and the Heisenberg term decoupled, the spinon Hamiltonian is quadratic,
\begin{equation}
H = \!\sum_{ij}\! \tilde{f}\dg_i \! \left[-\frac{xt_{ij}}{2}\tau_3 + U_{ij}\!\right]\!\tilde{f}_j + \frac{N\!\left[|\Delta_{ij}|^2+|\chi_{ij}|^2\right]}{J_{ij}},
\end{equation}
where we have introduced the $SU(2)$ matrix notation, 
\begin{equation}
U_{ij} = \zmatrix{-\chi_{ij}}{\Delta_{ij}}{\bar{\Delta}_{ij}}{\bar{\chi}_{ij}}.
\end{equation}
$\chi_{ij}$ generates a dispersion for the spinons, while $\Delta_{ij}$ pairs them.
The full Hamiltonian is given by $H + H_C$.
The physical electron, $c\dg \sim \langle b \rangle f\dg + \langle a \rangle f$ will either hop coherently, forming a Fermi liquid when $\Delta$ is zero, or will superconduct when $\Delta$ is nonzero.  The mean field phase diagram is obtained by minimizing the free energy with respect to these mean field parameters, $\chi_{ij}$ and $\Delta_{ij}$, 
\bea
\chi_{ij} & = & \frac{J_{ij}}{N}\langle f\dg_{i\alpha} f_{j\alpha}\rangle \cr
\Delta_{ij} & = & \frac{J_{ij}}{N}\langle \tilde{\alpha}f\dg_{i\alpha} f\dg_{j-\alpha}\rangle,
\eea
and enforcing the constraint on average, $\langle 
\sum_j f\dg_{j\alpha} f_{j\alpha} \rangle = \frac{N(1-x)}{2}$ and $\langle \sum_j\tilde{\alpha}f\dg_{j\alpha} f\dg_{j-\alpha} \rangle = 0$, where $\langle \cdots \rangle$ is the thermal expectation value.

The $J_1-J_2$ model will have two sets of bond variables, $\chi_\eta$ and $\Delta_\eta$, where $\eta$ indicates a link, $(ij)$. We assume that $\chi_1$ and $\chi_2$ are uniform, and allow $\Delta_1$ and $\Delta_2$ to be either s-wave  or d-wave.  When these order parameters are Fourier transformed, we find $\chi_\bk = \chi_1 \gamma_{1\bk} + \chi_2 \gamma_{2\bk} \equiv 2 \chi_1(c_x + c_y) + 4 \chi_2 c_x c_y$ and $\Delta_\bk = \sum_\eta \Delta_\eta \delta_{\eta \bk}$ is a combination of s-wave and d-wave pairing on the nearest and next nearest neighbor links,
\bea
\mathrm{extended}\;s & \; & 2\Delta_{1s} (c_x + c_y)\cr
d_{x^2-y^2}&\;& 2\Delta_{1d} (c_x - c_y) \cr
s_\pm&\;& 2\Delta_{2s} (c_{x+y} + c_{x-y}) = 4 \Delta_{2s} c_x c_y\cr
d_{xy}&\;& 2\Delta_{2d} (c_{x+y} - c_{x-y}) = -4 \Delta_{2d} s_x s_y 
\eea
and we define $c_\eta = \cos k_\eta a$, $s_\eta = \sin k_\eta a$.  The full Hamiltonian (including the constraint) has the form,
\bea
H & = & \sum_\bk \tilde{f}\dg_\bk \left(-\frac{x\epsilon_\bk}{2}+ U_\bk + \lambda_3 \tau_3 +\lambda_1 \tau_1  \right)\tilde{f}_\bk \cr
&& + \mathcal{N}_s\sum_\eta\frac{N}{J_{\eta}}\left(|\Delta_{\eta}|^2+|\chi_{\eta}|^2\right) - \frac{N\mathcal{N}_s x\lambda_3}{2} 
\eea
where $\epsilon_k$ is the Fourier transform of $t_{ij}$, $U_\bk$ is the Fourier transform of $U_{ij}$ and $\lambda_1 = \frac{1}{2}(\lambda_+ + \lambda_-)$. ($\lambda_2$ is unnecessary if $\Delta$ is real).  This Hamiltonian can be diagonalized, and the spinons integrated out to yield the free energy,
\bea
%[\chi_\eta,\Delta_\eta,\lambda_1,\lambda_3]
F & = & -2N T \sum_{\bk} \log 2 \cosh \frac{\beta \omega_{\bk}}{2}\cr
&& + \mathcal{N}_s\sum_\eta\frac{4 N}{J_{\eta}}\left(|\Delta_{\eta}|^2+|\chi_{\eta}|^2\right) - \frac{N\mathcal{N}_s x\lambda_3}{2},
\eea
where $\omega_\bk = \sqrt{\alpha_\bk^2 + \beta_\bk^2}$,
$\alpha_\bk = \lambda_3 -\frac{x\epsilon_\bk}{2}+ \chi_\bk,$ and $\beta_\bk = \lambda_1 + \Delta_\bk$.
Minimizing this free energy leads to the four mean field equations,
\bea
\label{MFEtJ}
\partial F/\partial \chi_\eta & = & \int_\bk \frac{\tanh \frac{\beta \omega_\bk}{2}}{2\omega_\bk} \alpha_\bk \gamma_{\eta \bk} - \frac{4}{J_\eta} = 0\cr
\partial F/\partial \Delta_\eta & = & \int_\bk \frac{\tanh \frac{\beta \omega_\bk}{2}}{2\omega_\bk} \beta_\bk \delta_{\eta \bk} - \frac{4}{J_\eta} = 0\cr
\partial F/\partial \lambda_3 & = & \int_\bk \frac{\tanh \frac{\beta \omega_\bk}{2}}{2\omega_\bk} \alpha_\bk - x/2 = 0\cr
\partial F/\partial \lambda_1 & = & \int_\bk \frac{\tanh \frac{\beta \omega_\bk}{2}}{2\omega_\bk} \beta_\bk  = 0.
\eea
The first three are identical to those for the $U(1)$ slave boson mean field theories\cite{kotliarliu88}, but the last enforces the absence of s-wave pairing.  $\lambda_1$ acts as a pair chemical potential adjusting the regions of negative and positive gap.

\section{Simple examples}
\label{examples}

\fg{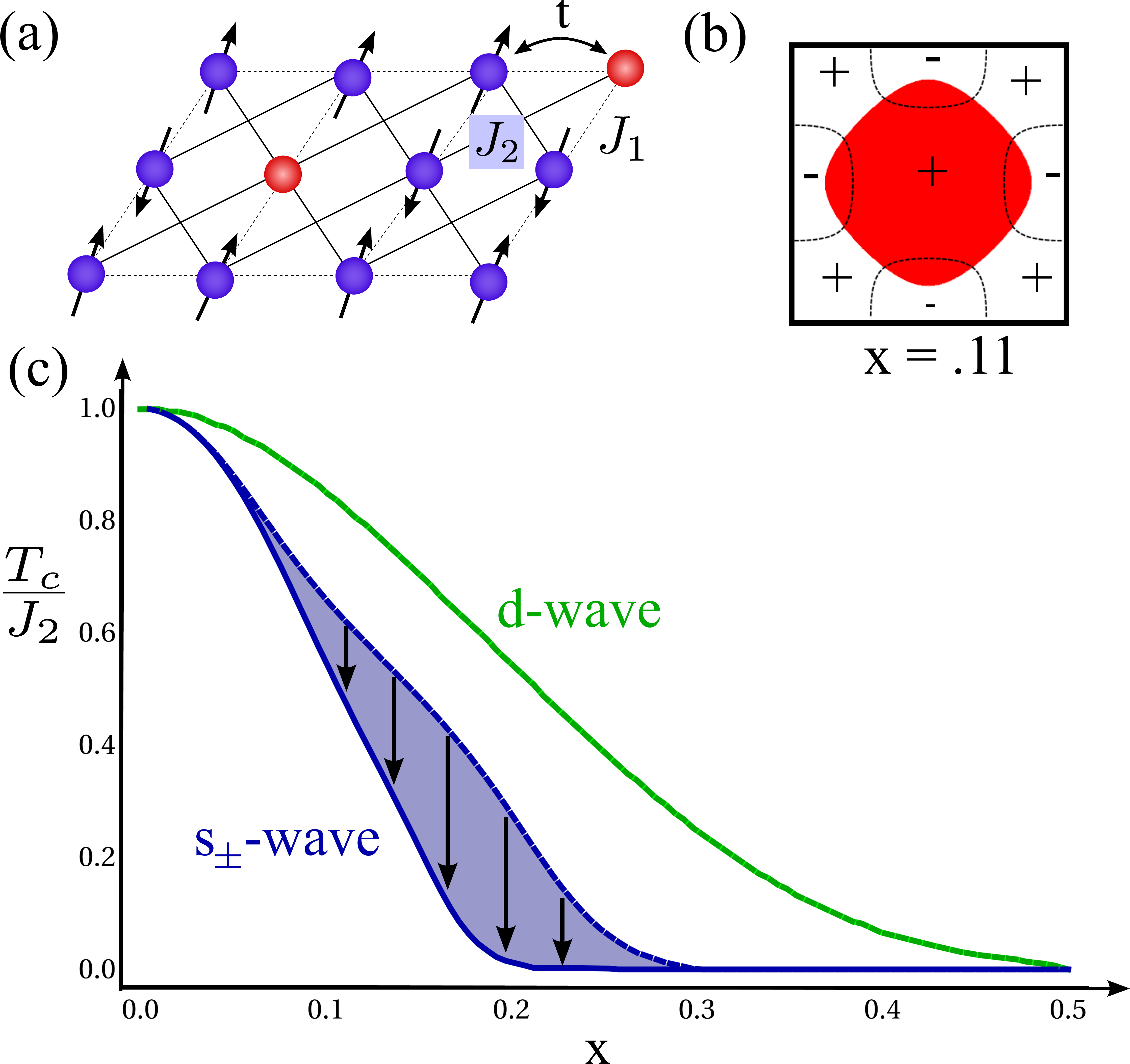}{t1J2}{(a) The $t_1-J_2$ model. (b) The Fermi
surface (holes shown in red) for the $t_1-J_2$ model at intermediate
doping.  In the superconducting state, the gap nodes follow the dashed
lines, separating regions of positive and negative gap.  (c) The
superconducting transition temperatures for the $t_1 -J_2$ model both
with (solid lines) and without the $\lambda_1$ constraint(dashed
lines), for s-wave (blue) and d-wave (green) superconductivity. d-wave
superconductivity is unaffected by the Coulomb repulsion, while the
s-wave transition temperature is decreased.}  

Now let us see this constraint in action, applied to several simple
cases.  First, we shall take the simplest lattice to exhibit $s_\pm$
pairing: the $t_1-J_2$ model shown in Fig. \ref{t1J2} (a).  Here, only
the next-nearest exchange coupling, $J_2$ and nearest neighbor
hopping, $t_1$ are nonzero, which leads to a single hole Fermi surface
with the potential for either $d_{xy}$ or $s_\pm$ pairing.  The
superconducting transition temperatures can be determined by setting
$\Delta_{2s/d} = 0^+$ and solving the mean field equations,
\eqref{MFEtJ} for $T_c$.  The results are shown in Fig.
\ref{t1J2}(c), where we have calculated the transition temperatures as
a function of doping, $x$ both with and without the $\lambda_1$
constraint.  The d-wave transition temperature is unaffected, as
$\partial F/\partial \lambda_1 = 0$ by symmetry, but the s-wave $T_c$
is suppressed.  Note that the two transition temperatures are
identical for $x = 0$.  Looking at the gap structure, Figure
\ref{t1J2}(b), we see that $\lambda_1$ has adjusted the gap nodes such
that there are equal amounts of positive and negative gap density of
states, eliminating the Coulomb repulsion.  As there is only one Fermi
surface in this example, there are necessarily line nodes even in the
s-wave state. The energetic advantage of a fully gapped s-wave
Fermi surface is thus lost, so that d-wave superconductivity, which
requires no costly adjustment of the nodes,  becomes energetically
favorable for this lattice.

\fg{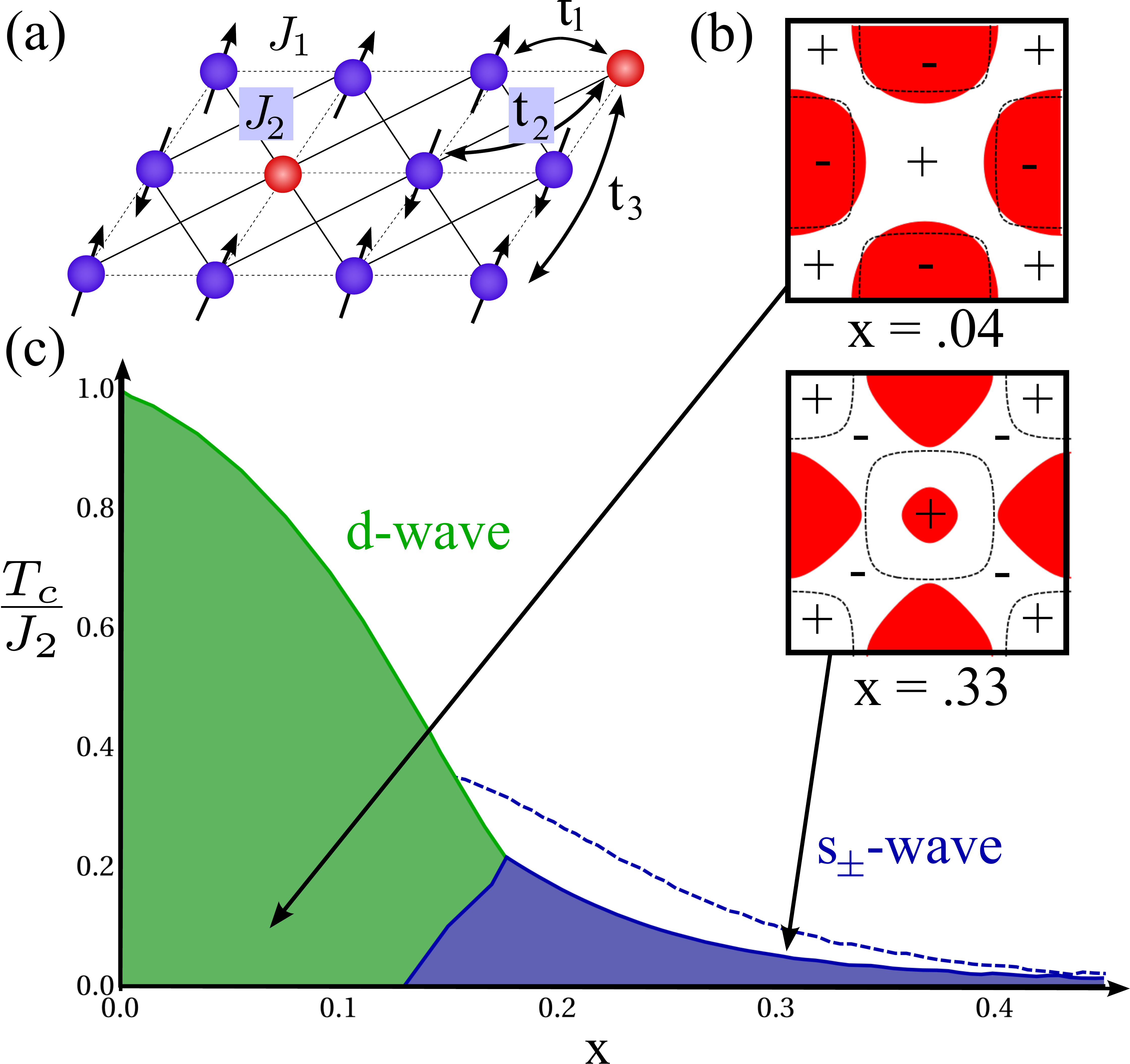}{tallJ2}{(a) The $t_1-t_2-t_3 -J_2$ model. (b)
The Fermi surface (holes shown in red) for the $t_1-t_2-t_3 -J_2$
model at both low (single Fermi pocket) and intermediate doping (two Fermi pockets).  The gap
nodes of the superconducting state are indicated with dashed lines.
(c) The doping phase diagram for the $t_1-t_2-t_3 -J_2$ model,
calculated with the $\lambda_1$ constraint (solid lines) and without (dashed lines).  There is a quantum phase
transition between d-wave pairing (green) and s-wave pairing (blue) as
doping increases and the s-wave states become fully gapped.}  

However, if there are multiple Fermi surfaces, $s_\pm$
superconductivity can gap out both surfaces with opposite signs.  If
we tune the $t_1-t_2-t_3$ hoppings, keeping only $J_2$, we can obtain
such a Fermi surface, with two hole pockets, as shown in Figure
\ref{tallJ2}(a,b).  Again, we calculate the s-wave and d-wave
transition temperatures in the presence of the pseudopotential terms,
showing the phase diagram in Figure \ref{tallJ2}(c).  
Our one-band approach makes this difficult, as the
size of the pockets shrinks with increasing doping.  The s-wave order
parameter has line nodes for low doping, which recede to point nodes
and then vanish as the Fermi surface becomes fully gapped at larger
dopings, where the s-wave superconductivity is more favorable than
d-wave, causing a d-wave to s-wave quantum phase transition as a function of doping.
If we had equally balanced hole and electron pockets at zero doping,
s-wave would likely win out over d-wave at all
dopings.

\section{Discussion}\label{discussion}

This study of the symplectic-$N$ t-J model illustrates the
importance of incorporating the Coulomb pseudopotential into any
strongly correlated treatment of $s_\pm$ superconductors.
The symplectic-$N$ scheme provides the first mean field solution of the $t-J$
model that is both controlled and superconducting. The large-$N$ limit
is identical to previous mean-field studies\cite{kotliarliu88}, but
contains the additional constraint fields $\lambda_{\pm}$ which
enforce the constraint $\vec{ \Psi }=0$.
For d-wave pairing, this
constraint is inert, as the s-wave component of the pairing is
zero by symmetry, but this constraint plays a very active role 
for s-wave pairing, acting
as a pair chemical potential that adjusts the gap nodes to eliminate
any on-site pairing.  As such, these models can capture the full
variety of gap physics proposed in the iron-based superconductors:
from line nodes to point nodes to two different full gaps that are not
otherwise expected in a local picture.  Properly accounting for the
adjustment of the line nodes is essential when comparing the relative energies
of d-wave and s-wave pairing states.

However, the large-$N$ limit suffers from an over-abundance of coherence, due to the ubiquity of the boson condensation.  As such, the only phases captured here are Fermi liquids and superconductors, and studying the effects of $1/N$ corrections is an important future direction.  This application is especially relevant to the cuprates, where there have been many intesting, but uncontrolled corrections to the mean field theories, revealing pseudogap-like phases formed by pre-formed pairs and incoherent metallic regions\cite{wenlee96,lee06}.  A controlled $1/N$ study of the phase diagram of the $t-J$ model studying the differences between s-wave and d-wave pairing should be of great interest.

While the $t-J$ models taken in this paper illustrate the basic effect of the Coulomb pseudopotential on strongly correlated superconductors, they are but poor approximations of the real materials, due to the single band approximation.  A better theory would involve multiple orbitals per site coupled by a ferromagnetic Hund's coupling, $-|J_H| \vec{S}_{\mu i} \cdot
\vec{S}_{\mu'i}$ between spins in different orbitals, $\mu \neq \mu'$ on
the same site.  Current large-$N$ techniques cannot treat such a ferromagnetic coupling, but future work might introduce an uncontrolled mean field parameter or take $J_H \rarrow \infty$, which may prove more tractable.

Interestingly, while the majority of the iron-based superconductors
have at least two electron and hole pockets, there are a handful of
``single band" materials: there are the end members
KFe$_2$As$_2$\cite{rotter08b} and
K$_{1-x}$Fe$_{2-y}$Se$_2$\cite{guo10}, which appear to have only
hole\cite{sato09} or electron pockets\cite{qian11}, respectively;{ and
the single layer FeSe, which has a single electron
pocket\cite{wang12,liu12}}.  In this local treatment, KFe$_2$As$_2$'s
single hole pocket must lead to a nodal d-wave
superconductor\cite{KFA}, as in the $t-J_2$ example above, where the
$s_\pm$ transition temperature is always smaller than the $d$-wave
temperature.  A d-wave gap is strongly suggested by recent heat
conductivity measurements\cite{reid12}.  On the other hand,
K$_{1-x}$Fe$_{2-y}$Se$_2$ { and single-layer FeSe} have electron pockets, which can develop node-less $d$-wave order, as originally discussed from the weak coupling approach\cite{mazin11,wang11,maier11}.  Including the Coulomb pseudopotential could again become important in this $d$-wave system if the tetragonal symmetry were broken.

Finally, an intriguing open problem in the iron-based superconductors is the relationship between the local quantum chemistry and the superconducting order\cite{ong11}.  The strong dependence of the superconducting transition temperature on the Fe-As angle\cite{dai08,yamada08} suggests that there might be a more local origin of superconductivity, similar to the composite pairs found in heavy fermion materials described by the two-channel Kondo lattice\cite{flint08}.  These two origins of $s_\pm$ pairing could then work in tandem to raise the superconducting transition temperature\cite{flint10}, and as such a future generalization of this work to take into account both the local iron chemistry and the staggered tetrahedral structure is highly desirable.  Such tandem pairing might explain the robustness of these superconductors to disorder on the magnetic iron site \cite{hhwen}.

Acknowledgments.
The authors wish to acknowledge discussions with Andriy Nevidomskyy, Rafael Fernandes and Valentin Stanev related
to this work. 
This work was supported by DOE grant DE-FG02-99ER45790.

{\section{Appendix}

In this section, we show that the operator combination
\begin{eqnarray}\label{l}
{\cal C}&=& %\frac{1}{2}S_{\alpha \beta }S_{\beta\alpha} - X_{\alpha 0}X_{0\alpha }+X_{0\alpha }X_{\alpha0}- (\tilde{X}_{00})^{2}\cr
 \frac{1}{2}S_{\alpha \beta }S_{\beta\alpha} +[X_{0\alpha },X_{\alpha
0}]- (\tilde{X}_{00})^{2}
\end{eqnarray}
commutes with the Hubbard operators, where $\tilde{X}_{00}= X_{00}+1$.
${\cal C}$ is therefore the quadratic casimir of the symplectic supergroup SP(N$\vert$1).
We also show that
\begin{equation}\label{}
C = (N/2)^{2}-1 - \vec{\Psi}_j^2
\end{equation}
in the symplectic slave boson representation.

The Hubbard operators $X_{0\alpha }$, $X_{\alpha 0}$ and $X_{00}$,
together with the symplectic spin operators, $S_{\alpha \beta }=
X_{\alpha \beta }-\frac{1}{N}X_{\lambda\lambda}$, 
form a closed superalgebra:
\begin{eqnarray}\label{l}
[S_{\alpha \beta },S_{\gamma \delta }]&=&  \delta_{\beta \gamma
}S_{\alpha \delta } - \delta_{\gamma\beta}S_{\alpha \delta} +\delta_{\bar \beta \delta} \tilde{\gamma} \tilde{\beta} S_{\alpha\bar\gamma} -\delta_{\gamma \bar \alpha} \tilde{\alpha}\tilde{\delta}S_{\bar \delta \beta}
\cr
[X_{0 \alpha },S_{\beta \gamma} ]&=&X_{ 0\gamma}\delta_{\alpha \beta }
+\tilde{\alpha}\tilde{\beta}X_{0\bar \beta }\delta_{\alpha \bar  \gamma}
\cr
\{X_{0\alpha },X_{\beta 0 }\}&=& S_{\beta \alpha }+ \delta_{\alpha \beta}\tilde{X}_{00}\cr
\{X_{0\alpha },X_{0\beta  }\}&=& 0
\cr
[X_{ 0\alpha },{X}_{00}]&=& -X_{0\alpha }.
\end{eqnarray}
Greek indices indicate spin indices $\alpha \in [\pm 1/2, \pm 3/2, \pm j]$ where $j=N/4$ and $N$ is even.  For simplicity, we use the notation $\bar \alpha = - \alpha$ and $\tilde{\alpha} = {\rm sgn}(\alpha)$. 
The operator 
$S_{\alpha \beta }= X_{\alpha \beta }-\frac{1}{N}\delta_{\alpha \beta
}X_{\lambda\lambda}$ is the traceless symplectic spin operator, while
the subsiduary operator,
$\tilde{X}_{00}= X_{00}- \frac{1}{N}\sum_{\alpha }X_{\alpha \alpha }$. 
This graded Lie algebra
defines the properties of the generators  of the 
symplectic supergroup SP(N$\vert$1).
This superalgebra is faithfully reproduced by the slave boson
representation
\begin{eqnarray}\label{l}
X_{\alpha \beta }&=& f\dg_{\alpha }f_{\beta }+\tilde{\alpha}\tilde{\beta}f_{\bar\alpha }
 f\dg_{\bar \beta }\cr
X_{\alpha 0}&=& f\dg_{\alpha }b+ \tilde{\alpha}f_{\bar \alpha }a\cr
X_{0\alpha }&=& b\dg  f_{\alpha }+ a\dg \tilde{\alpha}f_{\bar \alpha }\cr
X_{00}&=& b\dg b + a\dg a 
\end{eqnarray}
while the spin and subsiduary operator, $\tilde{X}_{00}$ are given by 
\begin{eqnarray}\label{l}
S_{\alpha \beta }&=& X_{\alpha \beta }-\frac{1}{N} \delta_{\alpha \beta
}X_{\lambda\lambda}= f\dg_{\alpha }f_{\beta }- \tilde{\alpha}\tilde{\beta}f\dg_{\bar \beta }f_{\bar
\alpha }\cr
\tilde{X}_{00}&=&
X_{00}+ \frac{1}{N}X_{\alpha \alpha }= b\dg  b+ a\dg a+ 1.
\end{eqnarray}

By inspection, 
${\cal C}$ contains only rotationally invariant combinations
of the Hubbard operators and each term 
leaves the number of slave bosons unchanged, so that it
commutes with $S_{\alpha \beta }$ and $X_{00}$. We now show by direct evaluation
that it also commutes with the
fermionic Hubbard operators $X_{0\alpha }$ and $X_{\alpha 0}$

First we evaluate the commutator between  $X_{\alpha 0}$ and 
the spin part of the Casimir, 
\begin{eqnarray}\label{l}
[X_{0\alpha},S_{\beta\gamma}S_{\gamma\beta }]&=&
[X_{0\alpha},S_{\beta\gamma}]S_{\gamma\beta }
+S_{\beta \gamma}[X_{0\alpha},S_{\gamma\beta }]\cr
%&=&[X_{0\gamma}\delta_{\alpha \beta }+ \delta_{\alpha \bar \gamma}{\rm
%sgn} (\alpha \beta )X_{0\bar  \beta }] S_{\gamma\beta }\cr 
%&+& S_{\beta\gamma }[X_{0\beta}\delta_{\alpha \gamma}+ \delta_{\alpha \bar \beta}{\rm
%sgn} (\alpha \gamma )X_{0\bar  \gamma }]\cr
&=&
 X_{0\gamma} S_{\gamma \alpha }+X_{0\bar \beta }S_{\bar \alpha \beta
}{\rm sgn } (\alpha \beta )\cr&
+& S_{\beta \alpha }X_{0\beta }+ {\rm sgn} (\alpha \gamma) 
S_{\bar  \alpha \gamma}X_{0\bar \gamma}
 \end{eqnarray}
Using the identity
$
S_{\alpha \beta }= - {\rm sgn} (\alpha \beta )S_{\bar  \beta\bar \alpha }
$,
we can convert this expression into the form
\begin{eqnarray}\label{a1}
[X_{0\alpha},S_{\beta\gamma}S_{\gamma\beta }]
%&=&  
%X_{0\gamma}
%S_{\gamma \alpha }-X_{0\bar \beta }S_{\bar  \beta \alpha 
%}{\rm sgn } (\alpha \beta ){\rm sgn} (\bar  \alpha \beta )
%\cr
%&+& S_{\beta \alpha }X_{0\beta }- {\rm sgn} (\alpha \gamma)   % I believe there's typos in here, did not fix
%{\rm sgn} (\bar  \alpha \gamma)
%S_{\bar \gamma \alpha }X_{0\bar \gamma}
%\cr
&=&X_{0\gamma}
S_{\gamma \alpha }
+X_{0\bar \beta }S_{\bar  \beta \alpha 
}
+ S_{\beta \alpha }X_{0\beta }
+S_{\bar \gamma \alpha }X_{0\bar \gamma}
\cr
%&=& 2 ( X_{0\gamma}
%S_{\gamma \alpha }+ S_{\beta \alpha }X_{0\beta })\cr 
&=& 2 \{X_{0\beta }, S_{\beta \alpha } \}.
\end{eqnarray}
%So that 
%\begin{equation}\label{a1}
%[X_{0\alpha},\frac{1}{2}S_{\beta\gamma}S_{\gamma\beta }]= \{X_{0\beta }, S_{\beta \alpha } \}.
%\end{equation}
Next we evaluate
\begin{eqnarray}\label{a2}
-[X_{0\alpha },
[X_{0\beta },X_{\beta 0}]
] &=& - X_{0\beta }\{X_{0\alpha
},X_{\beta 0} \} -\{X_{0\alpha},X_{\beta 0} \}X_{0\beta }\cr
%&=& - \{X_{0\beta },S_{\beta \alpha }+ \tilde{X}_{00}\delta_{\alpha
%\beta } \}\cr
&=& -\{
X_{0\beta },S_{\beta \alpha } \} - \{
X_{0\alpha },\tilde{X}_{00}\}.
\end{eqnarray}
Finally, 
\begin{eqnarray}\label{a3}
[X_{0\alpha },\tilde{X}_{00}^{2}]&=& - [X_{0\alpha },\tilde{X}_{00}]\tilde{X}_{00}-
\tilde{X}_{00} [X_{0\alpha },\tilde{X}_{00}]\cr
%&=& X_{0\alpha }\tilde{X}_{00} + \tilde{X}_{00}X_{0\alpha }\cr
&=& \{X_{0\alpha },\tilde{X}_{00} \}
\end{eqnarray}
Adding (\ref{a1}), (\ref{a2}) and (\ref{a3}) together gives
\begin{equation}\label{}
[X_{0\alpha },{\cal C}] = 0.
\end{equation}
Since $X_{\alpha 0}= X\dg _{0\alpha}$ and ${\cal C}= {\cal C}\dg $ is
Hermitian, it follows that 
$[X_{\alpha 0  },{\cal C}] = 0.$
Thus, ${\cal C}$ commutes with all Hubbard operators, and is thus a Casimir
of the supergroup SP(N$\vert$1) generated by the symplectic operators. 

To evaluate the Casimir, we insert the slave boson form of the Hubbard
operators.  First, evaluating the spin part, we obtain
\begin{eqnarray}\label{l}
S_{\alpha \beta }S_{\beta \alpha }%&=&\cr
%\left(
%f\dg_{\alpha }f_{\beta }\right.&-& \left. f\dg_{\bar \beta }f_{\bar \alpha }{\rm sgn} 
%(\alpha\beta )
%\right)
%\left( 
%f\dg_{\beta }f_{\alpha }-
% f\dg_{\bar \alpha }f_{\bar \beta }{ \sgn
%} (\alpha\beta ) 
%\right)\cr
&=& 2 [f\dg_{\alpha} f_{\beta} f\dg_{\beta}f_{\alpha }- f\dg_{\bar
\beta }
f_{\bar \alpha } f\dg_{\beta }f_{\alpha }{\rm sgn} (\alpha \beta )]\cr
&=& 2\left[f\dg_{\alpha } (N-f\dg_{\beta }f_{\beta })f_{\alpha }
+ f\dg_{\bar \beta }f\dg_{\beta }f_{\bar \alpha }f_{\alpha }{\rm sgn} (\alpha \beta ) \right]\cr
&=& 2 \left[ n_{f} (N+2-n_{f}) - 4 \Psi\dg_{f}\Psi_{f}
\right]
\end{eqnarray}
where   % could drop sgn(alpha) entirely
$\Psi_{f}\dg = \sum_{\alpha >0}\tilde{\alpha}f\dg_{\bar \alpha }f\dg _{\alpha } 
$, 
while
\bea
\left[X_{0\alpha }, X_{\alpha 0}\right] %& = & n_b(N-n_f)+n_a n_f -n_f(1+n_b)-(N-n_f)(1+n_a)\cr
& = & (N-2n_f)(n_b-n_a) -N.
\eea
Combining the various terms in the Casimir, we obtain
\begin{eqnarray}\label{l}
{\cal C}&=& n_{f} (N+2-n_{f})-(N-2n_f)(n_b-n_a) -N \cr
&-& 4 (\Psi \dg_{f}+ b\dg a)(\Psi _{f}+ a\dg b)+ 4 b\dg a a\dg  b- (\tilde{X}_{00})^{2}.
\end{eqnarray}
By regrouping terms, we obtain
\begin{eqnarray}\label{l}
{\cal C}&=&- (n_{f}+n_{b}-n_{a}-\frac{N}{2})^{2}+ (n_{b}-n_{a})^{2}+
\left(\frac{N}{2} \right)^{2}\cr
&+& 2 (n_{f}+ n_{b}-n_{a}-\frac{N}{2})- 2 (n_{b}-n_{a})\cr
&-& \Psi\dg \Psi +4 b\dg a a\dg  b -
(\tilde{X}_{00})^{2}.
\end{eqnarray}
We now introduce the triad of operators
\begin{eqnarray}\label{l}
\Psi_{3}&=& n_{f}+ n_{b}-n_{a}- \frac{N}{2}\cr
\Psi \dg &=&  \Psi\dg _{f}+ b\dg a\cr
\Psi &=& \Psi_{f}+a\dg  b
\end{eqnarray}
where $[\Psi\dg ,\Psi  ]= \Psi_{3}$.  Alternatively $\frac{1}{2} (\Psi_{1}+ i \Psi_{2})= \Psi\dg$ and 
$\frac{1}{2} (\Psi_{1}- i \Psi_{2})= \Psi$.
The Casimir can then be simplified to
\begin{eqnarray}\label{l}
{\cal C}
%&=& - (\Psi_{3})^{2}+ (n_{b}-n_{a})^{2}+ (N/2)^{2}+ 2
%\Psi_{3}\cr
%&-&2 (n_{b}-n_{a})- (\Psi_{1}^{2}+ \Psi_{2}^{2})- 2 \Psi_{3}
%\cr
&=& - (\Psi_{3})^{2}+ (n_{b}+n_{a}+1)^{2}+ (N/2)^{2}-1
\cr
&-& (\Psi_{1}^{2}+ \Psi_{2}^{2})- (n_{b}+n_{a}+1)^{2}
\end{eqnarray}
The terms involving $n_{b}$ and $n_{a}$ completely cancel out, leaving
${\cal C}= \left(\frac{N}{2} \right)^{2}-1- (\vec{ \Psi }^{2}).$}

\end{document}